\documentclass[a4paper,11pt]{article}
\usepackage{pos}
\usepackage{subfig}

\graphicspath{{figs/}}

\newcommand{\me}{\mbox{Mini-EUSO}}
\newcommand{\je}{\mbox{JEM-EUSO}}

\newcommand{\uhecrs}{\mbox{UHECRs}}

\title{Machine Learning for Mini-EUSO Telescope Data Analysis}
\ShortTitle{Machine Learning for Mini-EUSO}

\author*[a,b]{Mario Bertaina}
\author[c]{Mikhail Zotov}
\author[d,c]{Dmitry Anzhiganov}
\author[a,b,e]{Dario~Barghini}
\author[f]{Carl~Blaksley}
\author[a,b]{Antonio~Giulio~Coretti}
\author[d,c]{Aleksandr~Kryazhenkov}
\author[g]{Antonio~Montanaro}
\author[a]{Leonardo~Olivi}

\affiliation[a]{Dipartimento di Fisica, Universit\`a di Torino\\
	Torino, Italy}
\affiliation[b]{{INFN, Sezione di Torino\\ Torino, Italy}}
\affiliation[c]{Skobeltsyn Institute of Nuclear Physics, Lomonosov
	Moscow State University\\
	Moscow, Russia}
\affiliation[d]{Faculty of Computational Mathematics and Cybernetics,
        Lomonosov Moscow State University\\ Moscow, Russia}
\affiliation[e]{INAF, Osservatorio astrofisico di Torino\\
	Torino, Italy}
\affiliation[f]{RIKEN\\
	Wako, Japan}
\affiliation[g]{Department of Electronics and Telecommunications, Polytechnic University of Turin,\\ Corso Duca degli Abruzzi 24, 10129 Turin, Italy}

\onbehalf{for the JEM-EUSO collaboration} 

\emailAdd{bertaina@to.infn.it}

\abstract{Neural networks as well as other methods of machine learning
(ML) are known to be highly efficient in different classification tasks,
including classification of images and videos.  Mini-EUSO is a
wide-field-of-view imaging telescope that operates onboard the
International Space Station since 2019 collecting data on miscellaneous
processes that take place in the atmosphere of Earth in the UV range.
Here we briefly present our results on the development of ML-based
approaches for recognition and classification of track-like signals in
the Mini-EUSO data, among them meteors, space debris and signals the light curves and
kinematics of which are similar to those expected from extensive air
showers generated by ultra-high-energy cosmic rays. We show that even
simple neural networks demonstrate impressive
performance in solving these tasks.}

\ConferenceLogo{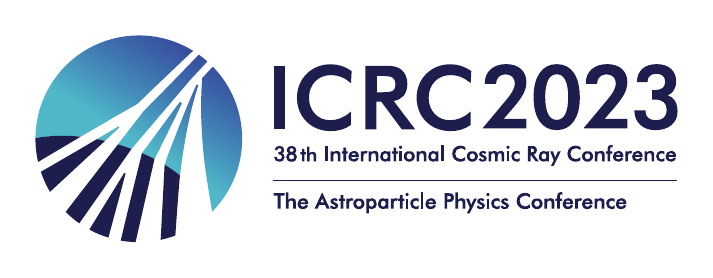}

\FullConference{%
38th International Cosmic Ray Conference (ICRC2023)\\
  26 July - 3 August, 2023\\
  Nagoya, Japan}

\begin{document}
\maketitle

\section{Introduction}

The \je{} (Joint Exploratory Missions for Extreme Universe Space Observatory) collaboration is developing a program of studying ultra-high energy cosmic rays (\uhecrs) with a wide angle telescope from a low Earth orbit~\cite{Bertaina:2021+i}.  The idea is based on the possibility to register fluorescence and Cherenkov radiation in the ultraviolet (UV) range that is emitted during development of extensive air showers generated by primary particles hitting the atmosphere.

It was clear from the early stages of the development of the program that an orbital instrument of this kind can also collect information about other processes taking place in the atmosphere in the UV~\cite{ABDELLAOUI2017245}. This was fully confirmed by the TUS mission, which took place on board the Lomonosov satellite in 2016--2017~\cite{AdvSR2022}. These days, the \me{} telescope is being operated on board International Space Station (ISS) as a part of an agreement between the Italian Space Agency (Agenzia Spaziale Italiana) and Roscosmos (Russia)  with the aim of studying transient atmospheric phenomena, meteors, searching for interstellar meteors and strange quark matter, and mapping nocturnal emission of the atmosphere in the UV~\cite{2023RSEnv.284k3336C}. Here we briefly present results of applying neural networks and other machine learning methods aimed at recognition and reconstruction of track-like signals in the \me{} data. These are tracks of meteors and space debris as well as signals of short flashes with light curves and kinematics similar to those expected from extensive air showers (EAS) of extreme energies. We shall call the later signals as EAS-like.


\section{Mini-EUSO Telescope}

The \me{} telescope is equipped with two Fresnel lenses with a diameter of 25~cm each, and a focal surface (FS) composed of $6\times6$ Hamamatsu R11265-M64 MAPMTs. Every MAPMT has 64 pixels, thus providing 2304 pixels in total. Each MAPMT has a BG3 UV-band glass filter. \me{} has a wide field of view (FoV) of~$44^\circ$$\times$$44^\circ$ with spatial resolution (FoV of one pixel) equal to $6.3~\text{km}\times6.3~\text{km}$.

\me{} collects data in three modes. The main mode of operation (D1) mode has a time resolution of 2.5~$\mu$s. This is called a D1 gate time unit (GTU). The next, D2 mode, records data integrated over 128 D1 GTUs. Finally, the D3 mode operates with data integrated over $128\times128$ D1 GTUs resulting in time resolution of 40.96~ms. The D1 and D2 modes have triggers, while data collected in the D3 mode is recorded as a stream, the duration of which corresponds to the time spent by the ISS at the nocturnal side of Earth. The orbital period of the ISS equals to 92.9 minutes, with nocturnal parts taking slightly longer than 1/3 of the period. Sessions of observations are performed approximately twice per month. A typical session includes eight subsets of data taken during nocturnal segments of orbits. A detailed description of the instrument can be found in~\cite{2023RSEnv.284k3336C}.

\section{Search for meteors and space debris}

\begin{figure}
    \centering
    \includegraphics[height=.2\textheight]{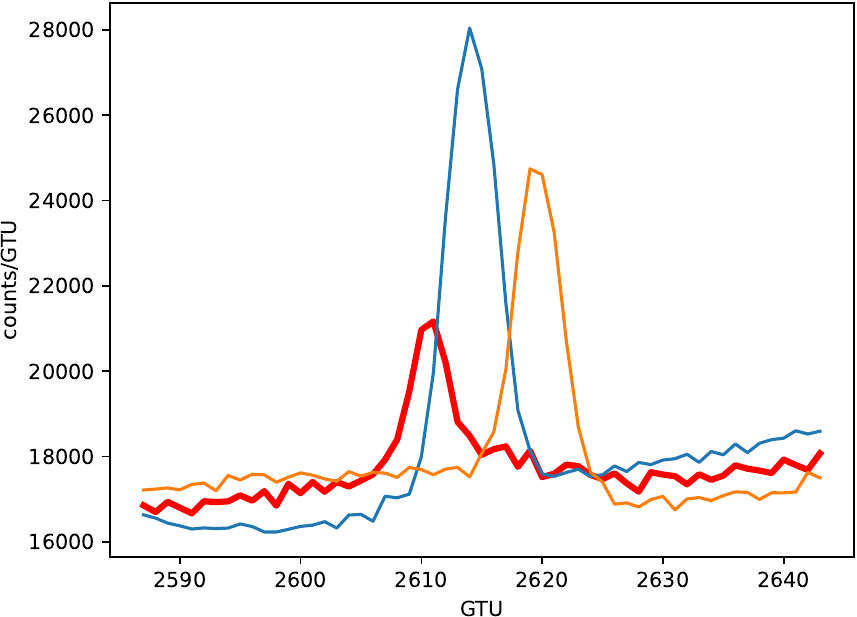}\quad\includegraphics[height=.22\textheight]{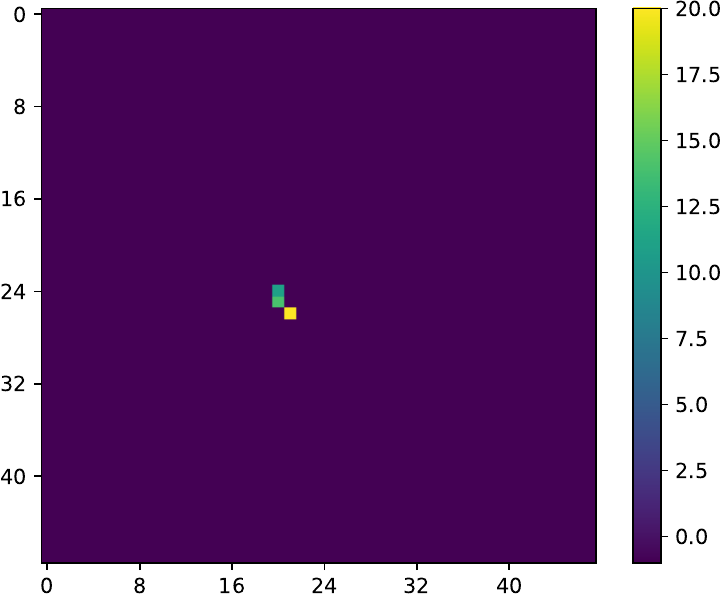}\\[2mm]
    \includegraphics[height=.22\textheight]{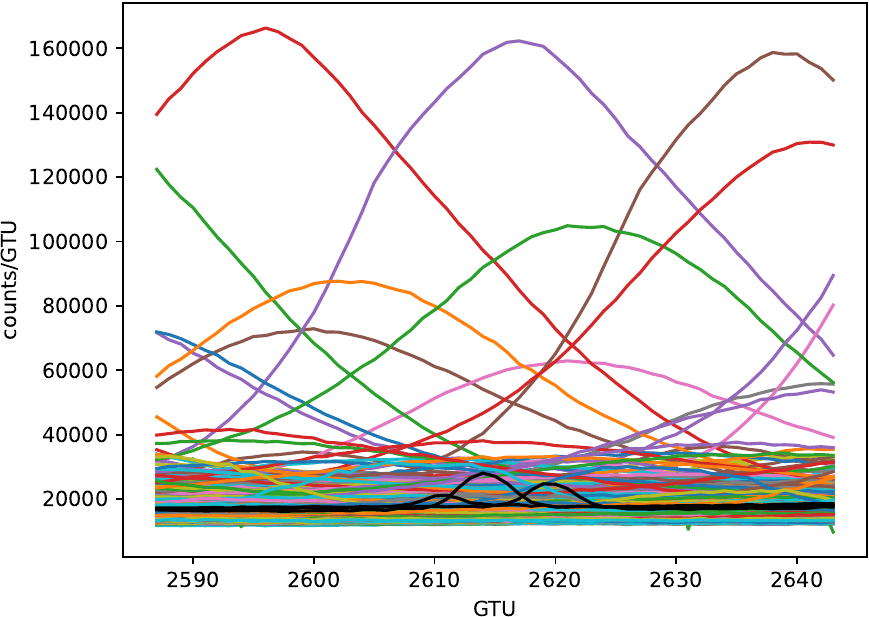}\quad\includegraphics[height=.22\textheight]{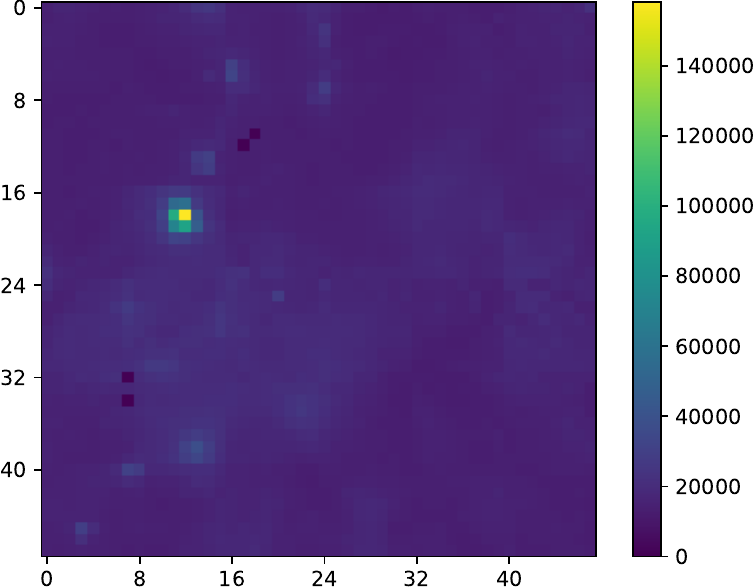}
    \caption{A typical meteor registered by \me. Top left: the meteor signal. Top right: location of the meteor track on the FS; colors indicate arrival time. Bottom left: the meteor signal (black curves) on the background of other signals registered simultaneously. Bottom right: a ``screenshot'' of the FS at the moment of the meteor signal peak.}
    \label{fig:meteor}
\end{figure}

Searches for meteors play an important role in the complementary \je{} program~\cite{ABDELLAOUI2017245}. However, recognizing a meteor track in the \me{} data is not as trivial as one may expect. Let us look at a typical meteor track registered by \me{} shown in Figure~\ref{fig:meteor}.

Figure~\ref{fig:meteor} illustrates several important features of meteor signals in the \me{} data:
\begin{itemize}\parskip=0pt
    \item these signals have a characteristic shape resembling the probability density function of a Gaussian distribution;
    \item they form quasi-linear tracks on the focal surface;
    \item the number of active pixels in meteor signals is small comparing to the complete focal surface;
    \item an amplitude of background (``non-meteor'') signals can be much greater than that of meteors.
\end{itemize}
Another important thing to mention is that observations are performed on a strongly varying background, which depends on the phase of the moon and the area in the FoV of \me{}.

We have developed several methods based on machine learning and neural networks to complement meteor studies based on the conventional approach.

\subsection{Stack-CNN for Space Debris and Meteor Detection and Tracking}

One of the new methods is based on a recent technique named \mbox{Stack-CNN}~\cite{2022JSSE....9...72M}, originally developed as an online trigger in a orbiting remediation system to detect space debris.

The Stack-CNN operational procedure can be divided in a stacking procedure used to enhance the signal over the background, and a convolutional neural network (CNN), acting as a binary classifier. The stacking method is used to generate a single image by summing over~$n$ frames that are shifted according to a combination of speed and direction of the space debris. If the combination matches the real one, the stacked image is enhanced by a factor of $\sqrt{n}$ in signal-to-noise ratio (SNR). 
Several combinations of speed and directions were tested and the CNN was used as a binary classifier to recognize a right combination from all the others. The method was tested and trained with space debris simulations using the \me{} framework. It showed extremely promising results, being able to detect signals with $\mathrm{SNR}\sim1\%$, which led to the exploration of offline applications. See \cite{2022JSSE....9...72M} for further details.

The Stack-CNN approach is extremely versatile as stacked combinations and the number of frames can be chosen depending on the physics of the object, while preserving the same trained neural network. Thus, the method was adapted to meteors as they share similar physical properties of space debris, such as apparent magnitude and speed.
The adapted algorithm was tested using real data from \me{} sessions. However, in $\sim28$~min of data files a total of 878 false positives were found. This was due to the extremely variable background illumination of \me, because as the telescope moves with the ISS, observations also include light emissions coming from cities, clouds and moon reflections. Thus, the Stack-CNN was reinforced with the Random Forest (RF) method acting as a binary classifier on meteor lightcurves. The dataset contains 1384 events from \me, equally distributed between meteors and the background. An advantage of using real data is that events include both Poissonian fluctuations and cities, which would have been difficult to simulate.
As a result, the trained architecture vastly outperformed the baseline Stack-CNN and standard approaches without machine learning techniques. In $\sim28$~min of data files, the algorithm found 64 more meteor events than the traditional trigger, with only 22 false positives being triggered.

The architecture was further validated using meteor simulations with different absolute magnitudes $M$. Out of 100 events of $M=+4$, 88 were detected by the algorithm, 11 more than within the conventional approach. The algorithm maintained a better performance even with fainter events of $M=+5$, detecting 70/100 events, 20 more than the standard approach. Finally, only 6/100 events of $M=+6$ were triggered, thus defining the detection limit of the model.

Simulations were used also to estimate the goodness of the speed and azimuth reconstruction of the Stack-CNN. It is worth noting that the triggered stacked speed and direction combination is apparent since Mini-EUSO comoves with the ISS. Thus, variables were converted to the real speed and azimuth values and the performance of the reconstruction was estimated using the average value and the standard deviation of the residual distribution. The residual distribution of the azimuth showed $\mu_\phi=(-3\pm4)^\circ$ and $\sigma_{\phi}=(46\pm3)^\circ$, whereas the speed residual distribution showed $\mu_v=(0\pm1)$~km/s and $\sigma_v=(17\pm1)$~km/s. The results indicate that there is no bias in the reconstruction, having average values compatible with~0. However, the standard deviations are larger than expected, which led to the exploration of other deep learning techniques for the track reconstruction. 

\subsection{Physics-Based Model for Meteor Dynamics}

The speed and direction reconstruction of meteors with the Stack-CNN algorithm is limited by the range and number of stacking combinations. Decreasing the former and increasing the latter could indeed yield a higher performance. However, this would not be a definitive solution and it would require more computational time. Thus, a new deep learning paradigm was investigated, involving implicit neural representation and meteor kinematics. 

\begin{figure}
    \centering
    \includegraphics[width=\textwidth]{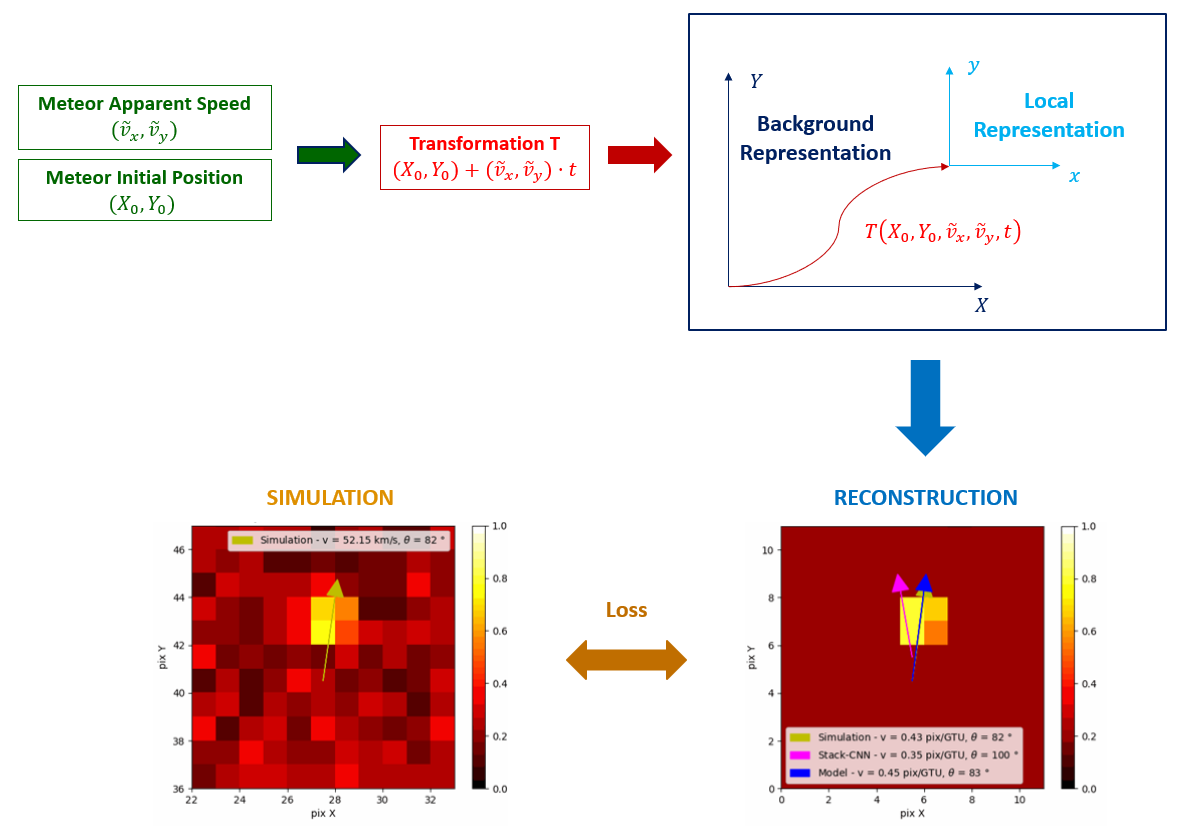}
    \caption{Concept of Physics-Model using implicit neural representation, inspired by~\cite{hofherr2023neural}}
    \label{fig:physmodel_arch}
\end{figure}

Implicit neural representations, also referred to as coordinate-based representations, are neural networks used to parameterize continuous and differentiable signals, such as images. In this way, the signal is encoded in the neural network parameters and it is often the only way an image could be parameterized, as an analytical function would be impossible to derive. 

In our work, the neural network performs a regression by mapping a pixel coordinate to the \me{} video sequences of meteor tracks. The video lasts 10~GTUs, and the resolution is $11\times11$ pixels, with the meteor track starting at the center.
Taking the architecture suggested in~\cite{hofherr2023neural} as inspiration, the physics dynamics has been implemented in the neural network with a time-spatial transformation~$T$ acting on two separate implicit neural representations.
The global pixel coordinates $(X,Y)$ are used as input for the first neural representation, returning the background photon counts.
Then, the pixel coordinates are mapped by function~$T$ to local reference frame $(x,y)$ of the meteor through its kinematics. The meteor starts moving in the first frame at pixel $(X_0,Y_0)$ at time $t = 0$. In the following frames, $t = 1,\dots, n_\mathrm{GTU}$, the meteor position is determined by its apparent speed~$\tilde{v}$~[pix/GTU] in x-axis and y-axis. 

\begin{equation*}
    \begin{cases}
        T(X_0,Y_0,\tilde{v},\theta,t) = (X_0 + \tilde{v}_x \cdot t, \ Y_0 + \tilde{v}_y \cdot t), \\
        (x,y) = (X,Y) - T(X_0,Y_0,\tilde{v}_x,\tilde{v}_y,t).
    \end{cases}
\end{equation*}

The local pixel depends on meteor dynamics and it is used as input in the second implicit neural representation, making the model based on the physics of the signal. The output of the second neural representation represents the signal itself and is summed over the background photon counts. A visual representation of the model is given in Figure~\ref{fig:physmodel_arch}. It is worth noticing also that in this case the selected dynamics is very simple, i.e., a point-like source moving with constant speed and direction, to start with the simplest case and to consider the only interesting parameters.

The architecture could be further refined by defining the transformation \textit{T} with the analytical solutions~\cite{met_ode} to the differential equations describing the physical problem of the meteor body deceleration in the atmosphere~\cite{met_phys}. Besides, the advantage of using implicit neural representation is that they are independent of the input resolution. Hence, pixel coordinates are not restricted to the $48\times48$ grid of \me, and the speed and direction reconstruction could be more accurate than other methods, such as the Stack-CNN.
The architecture we used is a multi-layer perceptron (MLP) with ReLU activation functions. It is worth noting, however, that such an architecture would underperform with RGB images or with higher resolution.

Other solutions would be mapping input coordinates to higher-dimensional inputs, called Fourier features \cite{tancik2020fourier}, or having an MLP with hidden sinusoidal activations. These architectures are called SIRENs, see~\cite{sitzmann2020implicit} for more details. Thus, with future work involving more complex dynamics and images at a higher resolution, these approaches could be much more efficient. 

\begin{figure}
    \centering
    \includegraphics[width=\textwidth]{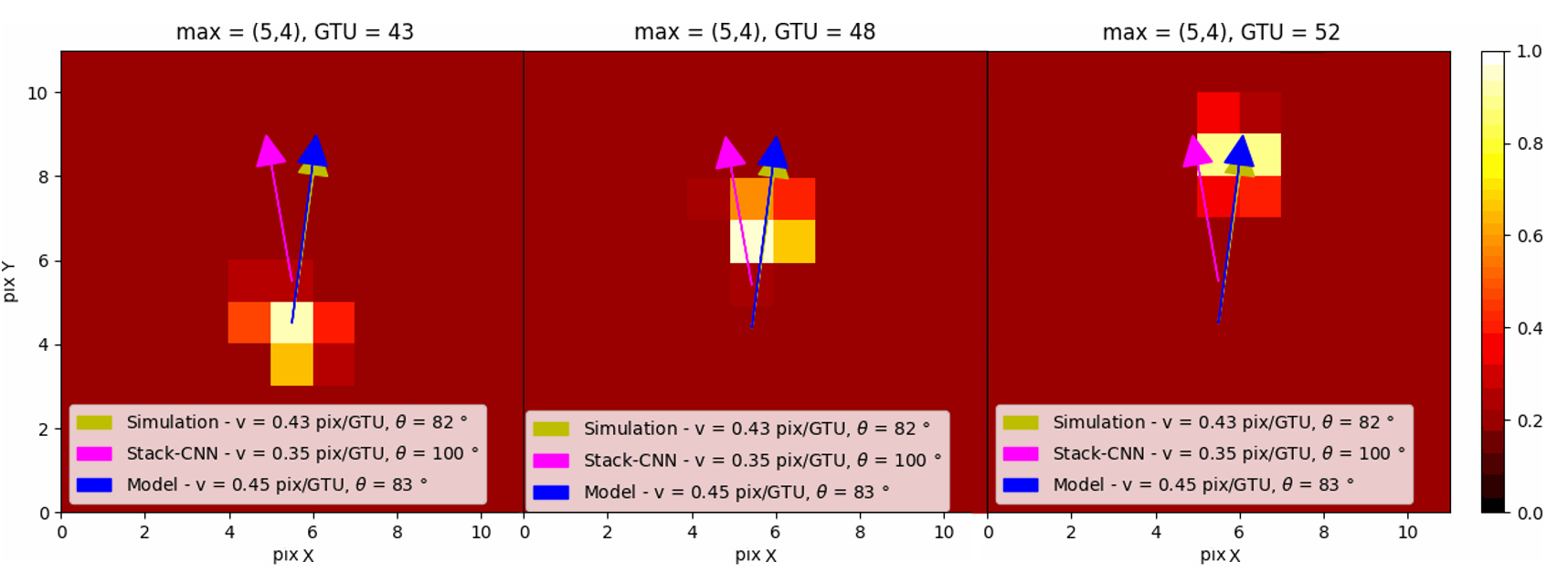}
    \caption{Reconstruction of Mini-EUSO meteor tracks. The comparison shows that the implicit neural representation (blue) is more similar to the simulation (yellow) than the Stack-CNN (pink)}
    \label{fig:rec_video}
\end{figure}

The model was trained and tested using the same meteor simulations presented for the Stack-CNN. Then, the reconstructed apparent variables were converted to the real speed and azimuth. The results outperformed the Stack-CNN, decreasing the resolutions of the residual distributions by $17^\circ$ for azimuth and 4~km/s for speed.
A visual comparison is given in Figure~\ref{fig:rec_video}, showing the new model (blue) outperforming the Stack-CNN (pink). In the future, this architecture could be further refined with a more complex meteor kinematics and tested using also real data. 

\subsection{Two-step Recognition of Meteor Tracks}

One more approach of recognizing meteor tracks in the \me{} data was implemented as a two-step procedure. At the first stage, a simple CNN was employed for binary classification of 3D chunks of data in the input stream into chunks that contain meteor tracks and all others (``non-meteors''). Input data was arranged in the form of $8\times8\times48$ packets, where $8\times8$ was the size of an area on the FS, and~48 was the number of time frames included in every packet. The packets were created using sliding windows to avoid signal loss. Training and testing the CNN was performed on a data set of 1068 meteors found with a conventional algorithm in the data of eight sessions that took place from 2019-11-19 till 2020-04-01. Training was performed on any seven sessions with the remaining session used for tests. The CNN demonstrated the highest performance on all possible combinations of input data and recognized properly all meteors.
Preliminary results of implementing this step were reported in~\cite{2022arXiv221203787Z} and will be presented in more detail elsewhere.

At the second stage, the task of image segmentation, i.e., recognition of hit pixels was applied to chunks containing meteor signals that were selected by the CNN. We tried several approaches: a MLP, a CNN, the RF, the K-nearest neighbours method (KNN), logistic regression and XGBoost. The two latter methods did not perform well, and we do not discuss them here. A comparison of results of the other four techniques is presented in Table~\ref{tab:stage2}.
\begin{table}
	\caption{The number of hit pixels with a meteor signal lost by different methods 
            at the second stage of meteor recognition. The first line indicates the session number.  The total number of hit meteor pixels in the respective data set is given in parentheses.}
        \centering
	\begin{tabular}{|l|c|c|c|c|c|c|c|c|}
		\hline
			 & 5     & 6      &7     & 8     & 11    & 12    & 13    & 14   \\
			 & (422) & (1428) &(80)  &(928)  & (958) & (492) & (457) & (630)\\
		\hline
		MLP & 2 & 2 & 0 & 7 & 5 & 0 & 0 & 2\\
		CNN & 8 & 1 &11 & 9 & 5 & 3 & 1 & 5\\
		RF  &34 &37 &20 &58 &53 &38 &28 &60\\
		KNN &34 &30 &23 &63 &50 &47 &19 &59\\
		\hline
	\end{tabular}
	\label{tab:stage2}
\end{table}
One can see that an MLP has outperformed all other tools demonstrating accuracy $>99\%$. The MLP had a very simple structure with two hidden layers with~96 and~64 neurons respectively. The CNN had one convolutional layer and three fully-connected layers. The RF and the KNN methods employed the amplitude and dispersion of the meteor peak as features (with a few other parameters tested).

\section{EAS-like events}

The above two-stage approach of recognizing meteor tracks was also tested for the search of EAS-like events in the D1 mode data collected with the time resolution of $2.5~\mu$s. The number of such events found in the \me{} data is too small to be used as a training data set. This made us use simulations. The combination of the CNN and the MLP described in the previous section demonstrated high performance though provided a big number of false positives when applied to the real data. This suggests that a more accurate simulation of the background is needed.

Another approach that we have tried can be called transfer learning. We have employed the fact that the shape and kinematics of meteor signals is very similar to those of EAS-like events even though they are evolving on totally different time scales. In order to use the ``meteor pipeline'', we changed the way of input data arrangement to take into account that D1 records contain 128 time frames. We trained the respective neural networks on D3 data packets of the size $8\times8\times64$. The trained pipeline was applied to the D1 records with every odd data frame omitted. Preliminary tests have demonstrated that the resulting performance of this ``transfered'' pipeline in terms of the number of found EAS-like events was comparable to that trained on simulated data. Results of the study will be reported elsewhere.

\acknowledgments

All neural networks discussed in the paper were implemented in Python with TensorFlow, PyTorch and scikit-learn software libraries.
The work of MZ, DA, and AK was funded by grant number 22-22-00367
of the Russian Science Foundation.

\bibliographystyle{JHEP}
\bibliography{ml4mini}
\newpage
{\Large\bf Full Authors list: The JEM-EUSO Collaboration\\}

\begin{sloppypar}
{\small \noindent
S.~Abe$^{ff}$, 
J.H.~Adams Jr.$^{ld}$, 
D.~Allard$^{cb}$,
P.~Alldredge$^{ld}$,
R.~Aloisio$^{ep}$,
L.~Anchordoqui$^{le}$,
A.~Anzalone$^{ed,eh}$, 
E.~Arnone$^{ek,el}$,
M.~Bagheri$^{lh}$,
B.~Baret$^{cb}$
% ...
}
\end{sloppypar}
\vspace*{.3cm}

{ \footnotesize
\noindent
% ...
$^{cb}$ Universit\'e de Paris, CNRS, AstroParticule et Cosmologie, F-75013 Paris, France\\
% ...
$^{ed}$ Istituto Nazionale di Fisica Nucleare - Sezione di Catania, Italy\\
% ...
$^{eh}$ INAF - Istituto di Astrofisica Spaziale e Fisica Cosmica di Palermo, Italy\\
% ...
$^{ek}$ Istituto Nazionale di Fisica Nucleare - Sezione di Torino, Italy\\
$^{el}$ Dipartimento di Fisica, Universit\`a di Torino, Italy\\
% ...
$^{ep}$ Gran Sasso Science Institute, L'Aquila, Italy\\
% ...
$^{ff}$ Nihon University Chiyoda, Tokyo, Japan\\ 
% ...
$^{ld}$ University of Alabama in Huntsville, Huntsville, AL, USA\\
$^{le}$ Lehman College, City University of New York (CUNY), NY, USA\\
% ...
$^{lh}$ Georgia Institute of Technology, USA\\
% ...
}

\end{document}